\documentclass[conference]{IEEEtran}
\IEEEoverridecommandlockouts
\usepackage{acronym} 
\usepackage{cite}
\usepackage{amsmath,amssymb,amsfonts}
\usepackage{algorithmic}
\usepackage{graphicx}
\usepackage{textcomp}
\usepackage{xcolor}
\usepackage{graphicx}  
\usepackage{caption} 
\usepackage{multirow}
\usepackage{array}
\usepackage{float}
\usepackage{placeins}
\def\BibTeX{{\rm B\kern-.05em{\sc i\kern-.025em b}\kern-.08em
    T\kern-.1667em\lower.7ex\hbox{E}\kern-.125emX}}
\usepackage{subcaption}
\usepackage{threeparttable} 
\usepackage[normalem]{ulem}

\usepackage[utf8]{inputenc} 
\usepackage[T1]{fontenc}    
\usepackage{url}

\newtheorem{theorem}{Theorem}

\newcommand{\BL}[1]{\textcolor{black}{#1}}

\acrodef{RTF}{relative transfer function}
\acrodef{GP}{Gaussian process}
\acrodef{GPR}{Gaussian process regression}
\acrodef{STFT}{short-time Fourier transform}
\acrodef{MMGP}{multiple-manifold GP}
\acrodef{CP}{conformal prediction}
\acrodef{LOO}{leave-one-out}
\acrodef{PI}{prediction interval}
\acrodef{GP}{Gaussian process}
\acrodef{STFT}{short-time Fourier transform}
\acrodef{ICP}{Inductive CP}
\acrodef{PSD}{power spectral density}
\acrodef{CPSD}{cross power spectral density}
\acrodef{TCP}{Transductive CP}
\acrodef{RT}{reverberation time}
\acrodef{UQ}{uncertainty quantification}
\acrodef{SSL}{Sound source localization}
\acrodef{ROI}{region of interest}
\acrodef{PSD}{power spectral density}
\acrodef{CPSD}{cross-PSD}

\begin{document}

\title{Conformal Prediction for Manifold-based Source Localization with Gaussian Processes\\
}

\author{
\IEEEauthorblockN{Vadim Rozenfeld and Bracha Laufer Goldshtein}
\IEEEauthorblockA{\textit{Department of Electrical Engineering}, \textit{Tel-Aviv University}, Tel-Aviv, Israel. \\
vadimroz@mail.tau.ac.il, blaufer@tauex.tau.ac.il}
}

\maketitle

\begin{abstract}
\BL{We address the problem of \ac{UQ} in the localization of a sound source within adverse acoustic environments. Estimating the position of the source is influenced by various factors, such as noise and reverberation, leading to
significant uncertainty. Quantifying this uncertainty is essential,
particularly when localization outcomes impact critical decision-making processes, such as in robot audition, where the accuracy of location estimates directly influences subsequent actions. 
Despite this, common localization methods offer point estimates without quantifying the estimation uncertainty. To address this, we employ \ac{CP}—a framework that delivers statistically valid \acp{PI} with finite-sample guarantees, independent of the data distribution. However, commonly used \ac{ICP} methods require a large amount of labeled data, which can be difficult to
obtain in the localization setting. To mitigate this limitation, we incorporate a semi-supervised manifold-based localization method using \ac{GPR}, with an efficient \ac{TCP} technique, specifically designed for \ac{GPR}. We demonstrate that our method generates statistically valid \acp{PI} across different acoustic conditions, while producing smaller intervals compared to baselines.\looseness=-1 } 
\end{abstract}


\section{Introduction}
\acf{SSL} is essential for various audio applications, including automated camera steering~\cite{huang2000passive}, speech enhancement and separation~\cite{mandel2010model}, and robot audition~\cite{rascon2017localization}. Numerous classical \ac{SSL} methods have been developed over the last decades, including Multiple Signal Classification (MUSIC)~\cite{schmidt1986multiple}, Estimation of Signal Parameters via Rotational Invariance Techniques (ESPRIT)~\cite{roy1989esprit}, Time-Difference-of-Arrival (TDOA)-based approaches~\cite{brandstein1997closed, schau1987passive}, Steered Response Power (SRP) techniques~\cite{dibiase2000high} and clustering of TDOA estimates across time-frequency bins~\cite{mandel2010model,schwartz2013speaker}. More recently, a variety of deep learning techniques have also been introduced~\cite{opochinsky2021deep, grinstein2023neural, grumiaux2022survey}. However, errors are unavoidable in any localization method and can significantly affect decision-making or other systems that rely on these outcomes. 
\BL{Robots, for instance, rely on precise location estimates to plan their actions effectively. When faced with high uncertainty, they can request additional guidance to enhance reliability and safety. Similarly, an automated camera might zoom out when location estimates are imprecise. These examples highlight the critical importance of accurately quantifying uncertainty in localization estimates.\looseness=-1 }

CP is a versatile framework for generating \acp{PI} with guaranteed coverage, ensuring that the true value falls within the \acp{PI} with a user-specified probability~\cite{vovk2005algorithmic}. Recently, \ac{CP} has been applied to source localization in a few studies. Previous work has explored its implementation with various uncertainty measures, such as Monte Carlo dropout, model ensembles, and quantile regression~\cite{khurjekar2023uncertainty}. In the context of multi-source direction-of-arrival (DOA) estimation, a Gaussian mixture model was employed to parameterize the conditional multi-source DOA distribution, and \ac{CP} was used to derive the \acp{PI} from the mixture model outputs~\cite{khurjekar2024multi}. However, these studies primarily focused on synthetic anechoic data and did not incorporate real speech signals. Additionally, they employ an \ac{ICP} approach, which relies heavily on a substantial amount of held-out calibration data in addition to the training data.\looseness=-1 

\BL{One significant challenge in source localization is the difficulty of obtaining large and diverse labeled data with known source positions.} 
Towards this end, we adopt a semi-supervised approach for source localization~\cite{laufer2016multiple,laufer2020data}. Our approach assumes access to a specific room where we can collect unlabeled data (i.e., measurements from unknown source locations) along with a small set of labeled samples with known source positions. To estimate the source positions, we employ \ac{GPR} using a specialized manifold-based kernel that captures the geometric structure of the acoustic features, leveraging both labeled and unlabeled data. We then integrate this method with a \ac{TCP} framework, which does not require additional held-out calibration data and can be efficiently applied to \ac{GPR}~\cite{papadopoulos2024guaranteed}.\looseness=-1

\section{Guaranteed coverage prediction intervals for Source Localization}
\subsection{Problem Formulation}\label{AA}
We consider a single source within a confined region surrounded by $M$ microphone nodes, each comprising two microphones. The source, positioned at $\mathbf{p} = [p_x, p_y]^T$, emits a signal $s(k, l)$, represented in the \ac{STFT} domain. Here, $ k \in \{1, \ldots, K\} $ denotes the frequency index and $ l \in \{1, \ldots, L\} $ denotes the frame index. The signal measured by the $ i $-th microphone of the $ m $-th node, is denoted by $ x_{i,m}(k, l) $, and can be expressed as:
\begin{equation}
x_{i,m}(k, l) = a_{i,m}(k;\mathbf{p}) s(k, l) + u_{i,m}(k, l),
\end{equation}
where $a_{i,m}(k;\mathbf{p})$ represents the acoustic transfer function between the source at position $\mathbf{p}$ and the $i$-th microphone of the $m$-th array, and $u_{i,m}(k, l)$ is an uncorrelated additive noise.\looseness=-1 

We extract an acoustic feature vector based on the \ac{RTF}, defined as $ h(k; \mathbf{p})=\frac{a_{2,m}(k; \mathbf{p})}{a_{1,m}(k; \mathbf{p})}$, which can be approximated by \BL{(noise term neglected)}:
\begin{equation}
\bar{h}_m(k)=\frac{\hat{S}_{m,21}(k)}{\hat{S}_{m,11}(k)}, 
\end{equation}
\BL{where, $\hat{S}_{m,11}(k)$ represents the \ac{PSD} of the first microphone, and $\hat{S}_{m,21}(k)$ denotes the \ac{CPSD} between microphones, both associated with the $m$-th node.} We focus on a pre-defined frequency range $k_1,\ldots,k_F$, where reliable \ac{RTF} estimates can be expected, and define the \ac{RTF} vector $\mathbf{h}^m=[\bar{h}_m(k_1),\ldots,\bar{h}_m(k_F)]^T$, associated with the $m$-th node.
We assume that the \ac{RTF} samples of the $m$-th node reside on a low dimensional manifold, denoted as $\mathcal{M}_m$. This assumption is based on the fact that in a given environment, where both the acoustic conditions and microphone positions remain relatively constant, the primary variable affecting the \ac{RTF} samples is the change in source position~\cite{laufer2015study, laufer2020data}. Since each node has a different perspective, it exhibits distinct relationships between \ac{RTF} samples, resulting in a unique manifold structure per node. Finally, we define an aggregated \ac{RTF} vector for all nodes by: 
\begin{equation}
\mathbf{h}=[\mathbf{h}^1,\ldots,\mathbf{h}^m]^T.
\end{equation}
\looseness=-1  

\BL{We assume access to a training} set of $n_L$ labeled samples associated with corresponding source positions
$\{\mathbf{h}_i, \mathbf{p}_i\}_{i=1}^{n_L}$, and $n_U$ unlabeled samples from unknown locations
$\{\mathbf{h}_j\}_{j=n_L+1}^{n}$, where $n=n_L+n_U$. Given a new test sample $\mathbf{h}_t$, \BL{ 
 we aim to localize} the source and produce a \ac{PI} $\Gamma^\delta(\mathbf{h}_t)$ such that:
\begin{equation}
\mathbb{P}\left(p_{t,c}\in\Gamma^\delta(\mathbf{h}_t)\right)\geq 1-\delta, 
\end{equation}
where $c\in\{x,y\}$ and $\delta$ is a user-defined confidence level.\looseness=-1  

\subsection{Multiple Manifold Gaussian Process}
To localize a source, we first define a mapping function associated with the $m$-th node, $f_m : \mathcal{M}_m \rightarrow \mathbb{R}$, which maps the \ac{RTF} sample $\mathbf{h}^{m} \in \mathcal{M}_m$ to the corresponding $x$ or $y$ coordinate of the source position. Let $f^m_{i} \equiv f^m(\mathbf{h}^m_{i})$ denote the position
evaluated by the function $f^m$ for the \ac{RTF} sample $\mathbf{h}^m_i$, where $i$ is a sample index. Note that the mapping is applied independently to each coordinate, with the coordinate index omitted for brevity. 
We assume that $f^m$ follows a zero-mean \ac{GP} with a manifold-based covariance function that quantifies the relation between samples by comparing their proximity to other samples on the manifold:\looseness=-1 
\begin{equation}
\textrm{cov}(\mathbf{h}^m_i, \mathbf{h}^m_j)\equiv\tilde{k}(\mathbf{h}^m_i, \mathbf{h}^m_j) =\sum_{r=1}^n  k^m(\mathbf{h}^m_{i}, \mathbf{h}^m_{r}) k^m(\mathbf{h}^m_{j}, \mathbf{h}^m_{r}),
\end{equation}
where $k^m : \mathcal{M}_m \times \mathcal{M}_m \rightarrow \mathbb{R}$ is a standard kernel function. We use a Gaussian kernel $k^m(\mathbf{h}^m_{i}, \mathbf{h}^m_{j}) = \exp \left( -\frac{\| \mathbf{h}^m_{i} - \mathbf{h}^m_{j} \|^2}{\sigma_m} \right)$, with a scaling factor $\sigma_m$. In addition, we assume that the per-node processes are jointly Gaussian, with covariance:\looseness=-1 
\begin{equation}
\textrm{cov}(\mathbf{h}^m_i, \mathbf{h}^q_j)\equiv\tilde{k}(\mathbf{h}^m_i, \mathbf{h}^q_r) =\sum_{r=1}^n  k^m(\mathbf{h}^m_{i}, \mathbf{h}^m_{r}) k^q(\mathbf{h}^q_{j}, \mathbf{h}^q_{r}).
\end{equation}

To fuse the information from the different nodes, we define a unified mapping function $f : \left(\bigcup_{m=1}^M \mathcal{M}_m \right) \rightarrow \mathbb{R}$, which associates an aggregated \ac{RTF} sample $\mathbf{h}_i$ with a specific coordinate of the source position $f_i \equiv f(\mathbf{h}_i)$. We define $f$ by the mean of the \acp{GP} of all nodes, forming a \ac{MMGP}. Thus, each position $f_i$ can be written as:\looseness=-1 
\begin{equation}
f_i = \frac{1}{M} \left( f^1_{i} + f^2_{i} + \cdots + f^M_{i} \right).
\end{equation}
Since the node-wise processes are jointly Gaussian, the combined process $f$ is also Gaussian with zero mean and covariance function
given by:
\begin{align}
\label{eq:combined_process}
\text{cov}(f_i, f_j) &\equiv \tilde{k}(\mathbf{h}_i, \mathbf{h}_j) = \frac{1}{M^2}\text{cov}\left(\sum_{m=1}^M f^m_i, \sum_{g=1}^M f^g_j\right) \nonumber \\
 &= \frac{1}{M^2}\sum_{m=1}^M\sum_{g=1}^M\text{cov}(f^m_i, f^g_j) \nonumber \\
&=\frac{1}{M^2} \sum_{r=1}^n \sum_{m=1}^M \sum_{ g=1}^Mk_m(\mathbf{h}^m_{i}, \mathbf{h}^m_{r}) k_g(\mathbf{h}^g_{j}, \mathbf{h}^g_{r}).   
\end{align}
This covariance function introduces a manifold-based kernel that incorporates the relations between all training points while integrating perspectives from different nodes.\looseness=-1 

We assume that the measured source positions satisfy a noisy observation model given by:
\begin{equation}
\label{eq:noisy_model}
p_i = f(\mathbf{h}_i) + \mathbf{\epsilon}_i, \quad i = 1, \ldots, n_L,
\end{equation}
where $\mathbf{\epsilon}_i \sim \mathcal{N}(0, \sigma_p^2)$ are i.i.d. Gaussian noise terms, independent of $f(\mathbf{h}_i)$. These noise terms account for uncertainties caused by imperfect measurements of the source positions during the acquisition of the labeled set. 
To localize a new test sample $\mathbf{h}_t$ from an unknown source position, we use an estimator based on the posterior probability $\mathbb{P}(f(\mathbf{h}_t) | \{\mathbf{h}_i, p_i\}_{i=1}^{n_L}, \{\mathbf{h}_i\}_{i=n_L+1}^{n})$, which is also Gaussian and is defined by the following mean and variance:\looseness=-1 
\begin{equation}
\hat{p}_t = \tilde{\mathbf{K}}_{L,t}^T \left( \tilde{\mathbf{K}}_L + \sigma_p^2 \mathbf{I}_{n_L} \right)^{-1} \mathbf{p}_L,
\label{eq:p_hat}
\end{equation}
\begin{equation}
\text{Var}(\hat{p}_t) = \tilde{K}_{tt} - \tilde{\mathbf{k}}_{L,t}^T \left( \tilde{\mathbf{K}}_L + \sigma_p^2 \mathbf{I}_{n_L} \right)^{-1} \tilde{\mathbf{k}}_{L,t},
\label{eq:p_var}
\end{equation}
where $\tilde{\mathbf{K}}_L$ is an $n_L \times n_L$ covariance matrix defined over the labeled samples, \BL{$\tilde{\mathbf{k}}_{L,t}$ is an $n_L \times 1$ covariance vector between the labeled samples and $f(\mathbf{h}_t)$, $\tilde{K}_{tt}$ is the variance of $f(\mathbf{h}_t)$, and $\mathbf{I}_{n_L}$ is the $n_L \times n_L$ identity matrix.} 
Note that although the variance in Eq.~\eqref{eq:p_var} reflects estimation uncertainty, if the model is not well-specified (e.g. wrong GP model hyperparameters or likelihood \cite{assael2014heteroscedastic, griffiths2021achieving}), these uncertainty estimates can become misleading. To mitigate this, we employ CP to produce valid \acp{PI}, even when the model is misspecified.\looseness=-1 

\begin{figure*}[!t]
    \centering
    \includegraphics[width=0.9\textwidth]{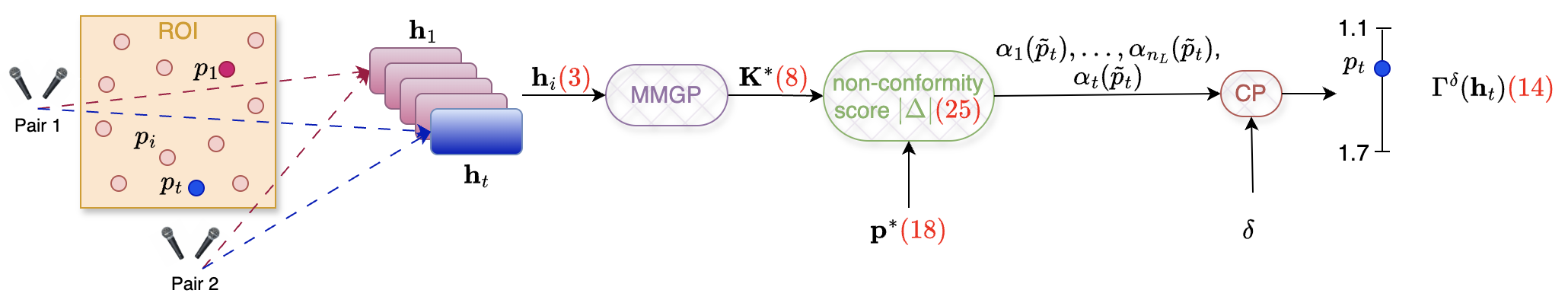}
    \vspace{-5.0pt}   
    \caption{\BL{\acs{MMGP} localization integrated with GPR-CP to derive the \ac{PI} $\Gamma^{\delta}(\mathbf{h}_t)$ for test position $p_t$, with relevant equations in red.}\looseness=-1}
\vspace{-15.0pt}    
    \label{fig:mmgp_cp_last}
\end{figure*}

\subsection{Conformal Prediction with Gaussian Process Regression}
\ac{CP} \cite{vovk2005algorithmic} is a framework that constructs PIs with a guaranteed coverage probability, defined by the user. 
There are two primary approaches to \ac{CP}: ICP\cite{papadopoulos2002inductive} and TCP\cite{Vovk2015TransductiveCP}. ICP enhances computational efficiency by dividing the dataset into separate training and calibration subsets, while TCP leverages the entire dataset for both tasks, resulting in higher statistical efficiency. However, in \ac{TCP}, the model must be re-computed for each test sample and every candidate label, which can be computationally expensive. Consequently, \ac{TCP} is typically applied to specific models~\cite{nouretdinov2001ridge, papadopoulos2011regression}, where this re-computation remains feasible. Recently, GPR-CP \cite{papadopoulos2024guaranteed} was introduced, combining \ac{GPR} with \ac{TCP} to generate \acp{PI} with guaranteed confidence levels while maintaining statistical efficiency. We will integrate this approach into the manifold-based \ac{GPR} localization method.\looseness=-1

\BL{The initial step in any \ac{CP} procedure involves defining a nonconformity score, which quantifies the extent to which a new data sample and its candidate label, \((\mathbf{h}_{t}, \tilde{p}_t)\), are atypical with respect to the existing dataset.}
A \ac{LOO} nonconformity score is defined as:
\begin{equation}
\label{eq:nonconformity_score}
    \alpha_i(\tilde{p}_t) = \lvert p_{i} - \hat{p}_{-i}(\tilde{p}_t)\rvert ,\quad i=1,\ldots, n_L,
\end{equation}
where $\hat{p}_{-i}(\tilde{p}_t)$ represents the estimator for the $i$-th sample given all other labeled samples, as well as the test sample, associated with a candidate label $\tilde{p}_t$. We will define $\hat{p}_{-i}$ more formally later in the context of \ac{GPR}. Similarly, the nonconformity score of the test sample is $\alpha_t(\tilde{p}_t) = \lvert \tilde{p}_t - \hat{p}_{t}\rvert$, where $\hat{p}_t$ is computed using Eq.~\eqref{eq:p_hat}, based on all labeled samples.\looseness=-1 

To statistically quantify how unusual the candidate label $\tilde{p}_t$ is, \ac{CP} computes a valid p-value for $\tilde{p}_t$ by comparing its nonconformity score to those of the existing data samples:
\begin{equation}
\label{eq:p_value}
    \mathbb{P}_\textrm{val}\left(\tilde{p}_t\right) = \frac{\lvert\{i = 1, \dots, n_L : \alpha_i(\tilde{p}_t) \geq \alpha_t(\tilde{p}_t)\}\rvert+1}{n_L+1}.
\end{equation}
Accordingly, we construct a \ac{PI} $\Gamma_{t}^\delta(\mathbf{h}_{t})$ as:
\begin{equation}
\label{eq:predictive_interval}
    \Gamma_{t}^\delta(\mathbf{h}_{t}) = \left\{ \tilde{p}_t :  \mathbb{P}_\textrm{val}\left(\tilde{p}_t\right) > \delta\right\},
\end{equation}
with valid finite-sample coverage, as formalized in the following theorem.\looseness=-1 
\begin{theorem}
(Marginal \ac{CP} coverage guarantee~\cite{vovk2005algorithmic}).
Suppose $(\mathbf{h}_1, p_1), \ldots ,(\mathbf{h}_{n_L}, p_{n_L}), (\mathbf{h}_{t}, p_{t})$ are exchangeable, then
the \ac{PI}~\eqref{eq:predictive_interval} satisfies:\looseness=-1 
\begin{equation}
    \mathbb{P}\left(p_t \in \Gamma_{t}^\delta(\mathbf{h}_{t})\right) \geq 1 - \delta.
\end{equation}
\end{theorem}


\BL{Note that the coverage property is marginal, holding on average over the distribution of the training set and the test point.} 
In principal, the computation of~\eqref{eq:nonconformity_score} requires re-estimating the model for each held-out combination and every candidate test position. Below, we outline how the \acp{PI} can be efficiently computed with a \ac{GPR} as the underlying model~\cite{papadopoulos2024guaranteed}. We utilize the fact that the \ac{LOO} estimator for \ac{GPR} can be written as
 \cite{rasmussen2006gaussian}:
\begin{equation}
\label{eq:mu_loo}
    \hat{p}_{-i} = p_i - \frac{[ (\mathbf{K}^* + \sigma_p^2 \mathbf{I}_{n_L+1})^{-1}\mathbf{p}^*]_i}{[(\mathbf{K}^* + \sigma_p^2 \mathbf{I}_{n_L+1})^{-1}]_{ii}},
\end{equation}
\begin{equation}
    \text{Var}(\hat{p}_{-i})= \frac{1}{[(\mathbf{K}^* + \sigma_p^2 \mathbf{I}_{n_L+1})^{-1}]_{ii}},
\label{eq:loo_variance}
\end{equation}
with $\mathbf{K}^*\in\mathbb{R}^{(n_L+1)\times (n_L+1)}$ formed by the kernel $\tilde{k}(\mathbf{h}_i,\mathbf{h}_j)$ in~\eqref{eq:combined_process} evaluated over
$\mathbf{h}_j, \mathbf{h}_j \in \{\mathbf{h}_i\}_{i=1}^{n_L} \cup \mathbf{h}_t$, and 
\begin{equation}
    \mathbf{p}^*=[p_1, \ldots, p_{n_L}, \tilde{p}_t]^T.
\end{equation}
Defining $\hat{\mathbf{p}}_\textrm{LOO}=[\hat{p}_{-1}, \ldots, \hat{p}_{-n_L}, \hat{p}_{t}]^T$, the vector of nonconformity scores $\Delta\equiv[\alpha_1(\tilde{p}_t),\ldots, \alpha_{n_L}(\tilde{p}_t), \alpha_t(\tilde{p}_t)]^T=\lvert \mathbf{p}^*-\hat{\mathbf{p}}_\textrm{LOO}\rvert$ is computed as:
\begin{equation}
\label{eq:papadopolous_nonconformity}
    \lvert \Delta\rvert=\big \lvert (\mathbf{K}^* + \sigma_p^2 \mathbf{I}_{n_L+1})^{-1}\mathbf{p}^* \,./ \text{diag}((\mathbf{K}^*+ \sigma_p^2 \mathbf{I}_{n_L+1})^{-1}) \big \rvert,
\end{equation}
where $./$ denotes element-wise division. We decompose the vector $\mathbf{p}^*$ as $\mathbf{p}^* = \mathbf{p}_a + \mathbf{p}_b= [p_1, \ldots, p_{n_L}, 0]^T+[0, \ldots, 0, \hat{p}_t]^T$. Thus, \eqref{eq:papadopolous_nonconformity} takes the form $\lvert\mathbf{a}+\tilde{p}_t\cdot \mathbf{b}\rvert$ where:
\begin{equation}
    \mathbf{a}=(\mathbf{K}^* + \sigma_p^2 \mathbf{I}_{n_L+1})^{-1}\mathbf{p}_a \,./ \text{diag}((\mathbf{K}^* + \sigma_p^2 \mathbf{I}_{n_L+1})^{-1}),
\end{equation}
\begin{equation}
    \mathbf{b}=(\mathbf{K}^* + \sigma_p^2 \mathbf{I}_{n_L+1})^{-1}\mathbf{p}_b \,./ \text{diag}((\mathbf{K}^*+ \sigma_p^2 \mathbf{I}_{n_L+1})^{-1}).
\end{equation}
Hence, the nonconformity scores $\lvert \Delta\rvert=\lvert \mathbf{a}+\tilde{p}_t\cdot \mathbf{b}\rvert$ are piecewise linear with respect to $\tilde{p}_t$ for each $\Delta_i$. Furthermore, since $\mathbb{P}_\textrm{val}(\tilde{p}_t)$ changes only when 
$\alpha_i(\tilde{p}_t)=\alpha_t(\tilde{p}_t)$, 
it eliminates the need to evaluate infinitely many candidates, thereby resulting in a feasible prediction algorithm. Specifically, for computing $\mathbb{P}_\textrm{val}(\tilde{p}_t)$, 
 we construct finite-sets for every labeled sample: 
\begin{align}
    S_i &= \left\{\tilde{p}_t : \alpha_i(\tilde{p}_t) \geq \alpha_{t}(\tilde{p}_t)\right\} \>\>\>\>\>\>\>\>  , i = 1, \dots, n_L\nonumber \\
    &=
    \left\{\tilde{p}_t : |a_i + \tilde{p}_t \cdot b_i| \geq |a_{n_L+1} + \tilde{p}_t \cdot b_{n_L+1} |\right\}.
\end{align}
Then, the p-value~\eqref{eq:p_value} takes the form:
\begin{equation}
   \mathbb{P}_\textrm{val}(\tilde{p}_t) = \frac{\lvert \{i = 1, \dots, n_L : \tilde{p}_t \in S_i\}\rvert + 1}{n_L + 1},
\end{equation}
and the interval $\Gamma_{t}^\delta(\mathbf{h}_{t})$ contains all candidates for which $\mathbb{P}_\textrm{val}(\tilde{p}_t)>\delta$, as defined in \eqref{eq:predictive_interval}. An efficient algorithm for finding all relevant intervals appears in~\cite{papadopoulos2024guaranteed, nouretdinov2001ridge}.\looseness=-1 


\begin{table*}[!t]
\centering
\caption{\BL{X-coordinate results. Lowest \ac{PI} widths are marked in boldface, and miscoverage instances are marked in red.}}
\vspace{-0.25em}
\begin{threeparttable}
\begin{tabular}{c c |cccc cc |c cccc ccc}
\hline
\multirow{1}{*}{$T_{60}$} & \multicolumn{7}{c}{$\hspace{7em}300$ ms} & \multicolumn{6}{c}{\hspace{1em}$700$ ms} \\
\hline
\multirow{2}{*}{SNR} & \multirow{2}{*}{Method} & \multicolumn{3}{c}{Coverage ($\%$)} & \multicolumn{3}{c|}{Mean PI [m]} & \multicolumn{3}{c}{Coverage ($\%$)} & \multicolumn{3}{c}{Mean PI [m]} \\
 &  & 90 & 95 & 99 & 90 & 95 & 99 & 90 & 95 & 99 & 90 & 95 & 99 \\
\hline
\multirow{3}{*}{\centering $5$~dB}
& GPR & 0.907 & {\color{red}0.938} & {\color{red}0.974} & 0.42 & 0.50 & 0.64 
& {\color{red}0.866} & {\color{red}0.910} & {\color{red}0.952} & 0.52 & 0.61 & 0.78\\
& Jackknife+ & 0.905 & 0.965 & 0.995 & \textbf{0.36} & 0.45 & 0.62 & 0.894 & 0.945 & 0.983 & 0.49 & \textbf{0.58} & 0.75  \\
& GPR-CP & 0.896 & 0.952 & 0.991 & \textbf{0.36} & \textbf{0.44} & \textbf{0.60} & 0.891 & 0.943 & 0.980 & \textbf{0.48} & \textbf{0.58} & \textbf{0.73} \\
\hline
\multirow{3}{*}{\centering $15$~dB}
& GPR & 0.908 & 0.949 & 0.982 & 0.35 & \textbf{0.41} & 0.53 & 0.915 & 0.951 & 0.981 & 0.45 & 0.53 & 0.69  \\
& Jackknife+  & 0.925 & 0.963 & 0.989 & 0.34 & 0.42 & 0.53 & 0.940 & 0.972 & 0.991 & 0.45 & 0.54 & 0.66\\
& GPR-CP & 0.909 & 0.953 & 0.988 & \textbf{0.33} & \textbf{0.41} & \textbf{0.50} & 0.925 & 0.951 & 0.988 & \textbf{0.44} & \textbf{0.51} & \textbf{0.63} \\
\hline
\end{tabular}
\end{threeparttable}
\label{tb:results_combined}
\end{table*}

In addition, the nonconformity scores can be normalized to obtain more precise \acp{PI}, which are tighter for inputs that are easier to predict and wider for
inputs that are more difficult to predict~\cite{papadopoulos2024guaranteed}.
We use the \ac{LOO}-variance~\eqref{eq:loo_variance} for normalization:
\begin{equation}
\label{eq:gamma_nonconformity}
\alpha_{i}(\tilde{p}_{t}) = \left\lvert \frac{p_i - \hat{p}_{-i}}{\sqrt[\gamma]{\text{Var}(\hat{p}_{-i})}} \right\rvert = \left\lvert \frac{p_i - \hat{p}_{-i}}{[(\mathbf{K}^* + \sigma_p^2 \mathbf{I}_{n_L+1})^{-1}]_{ii}^{-\frac{1}{\gamma}}} \right\rvert,
\end{equation}
where $\gamma$ controls the sensitivity to changes in the variance. Accordingly, the nonconformity scores \eqref{eq:papadopolous_nonconformity} take the form:
\begin{equation}
     \lvert \Delta\rvert=\big \lvert (\mathbf{K}^* + \sigma_p^2 \mathbf{I}_{n_L+1})^{-1}\mathbf{p}^* \,./ \text{diag}((\mathbf{K}^*+ \sigma_p^2 \mathbf{I}_{n_L+1})^{-1})^{1-\frac{1}{\gamma}} \big \rvert.
\end{equation}\looseness=-1 
An overview of the full process is provided in Fig.~\ref{fig:mmgp_cp_last}. \looseness=-1 

\section{Experiments and Results}\label{FAT}

\subsection{Experimental settings}\label{FAT}

\begin{figure}[t]
\vspace{-1.6em}
    \centering
    \includegraphics[width=0.48\linewidth]{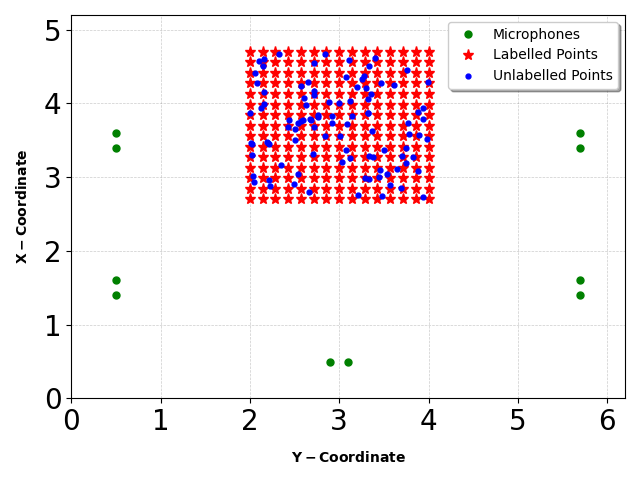}  
\vspace{-0.25em}
    \caption{The simulated room setup.}
    \label{fig:room}
\vspace{-1.5em}
\end{figure}

We simulated $5.2~\textrm{m}~\times6.2 \textrm{m}~\times3.5$m rooms with \acp{RT} of $300$~ms and $700$~ms\BL{~\cite{habets2006rirgenerator}}. Five ($M=5$) microphone pairs were placed around the room's walls. The source positions were confined to a $2~\textrm{m} \times 2$~m \ac{ROI} (see Fig.~\ref{fig:room}).
We generated $n_L = 225$ labeled samples, equally spaced with $0.133$~m resolution, $n_U = 100$ unlabeled samples and $n_T = 200$ test samples, both randomly sampled within the same region. 
Throughout our simulations, we used speech utterances from the LibriSpeech dev-set~\cite{panayotov2015librispeech}.
White Gaussian noise was added to all microphone signals \BL{with two SNR levels: 5 dB and 15 dB}. The signals were analyzed using an STFT with $1024$ frequency bins and $75\%$ frame overlap. The frequency range $150-1500$~Hz was used in the construction of the \ac{RTF} vectors. 
We set $\gamma=32$ for the score normalization~\eqref{eq:gamma_nonconformity}, as it provided the tightest \acp{PI}.\looseness=-1 

\subsection{Evaluation}
We compare the performance of GPR-CP against the standard \ac{GPR}, where \acp{PI} are derived directly from the estimated variance~\eqref{eq:p_var}. \BL{Additionally, we use the Jackknife+ algorithm~\cite{barber2021predictive} as another baseline that is similar in principal to CP. Jackknife+ constructs \acp{PI} based on the quantiles of \ac{LOO} predictions and their residuals, ensuring at least $1-2\delta$ coverage under data exchangeability assumption, which in practice is close to $1-\delta$ coverage.} 
We examine three coverage levels: $90\%$ ($\delta = 0.1$), $95\%$ ($\delta = 0.05$), and $99\%$ ($\delta = 0.01$). We assess the tightness of the \acp{PI} by their mean width, as narrower \acp{PI} are preferred, provided they meet the required coverage levels. Each experiment is repeated 10 times with different random unlabeled and test positions, and the average results across these trials are reported.\looseness=-1 

\subsection{Results}
The empirical coverage and mean \ac{PI} widths for all methods are summarized in Table~\ref{tb:results_combined}. Due to space constraints, we present the results for the X-axis only; similar trends were observed for the Y-axis. \BL{We observe that wider \acp{PI} are obtained for either higher reverberation or lower SNR levels. Noticeably, GPR-CP achieves the target coverage across all conditions, while obtaining the lowest \ac{PI} widths compared to the baselines.}
In contrast, GPR exhibits undercoverage at $5$~dB SNR, consistent with findings in~\cite{papadopoulos2024guaranteed}, while Jackknife+ tends to be conservative at $15$ dB SNR, resulting in slightly wider \acp{PI} compared to GPR-CP.\looseness=-1 

\begin{figure}[!t]
\vspace{-2.2em}
\begin{center}
\begin{subfigure}[b]{0.25\textwidth}
 \centering
 \includegraphics[width=\textwidth]{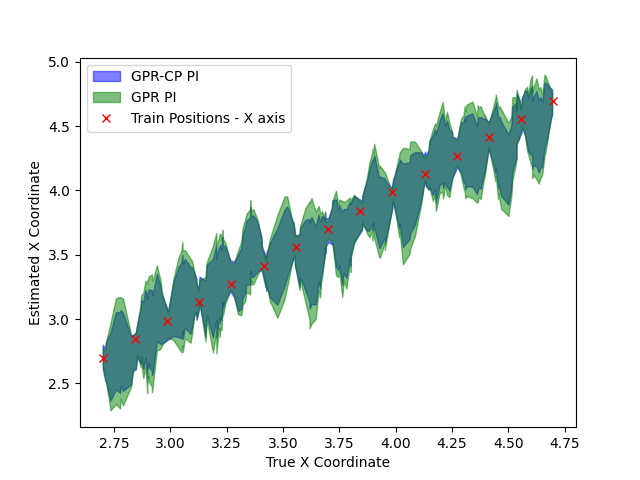}
\vspace{-2em}
\end{subfigure}
\vspace{-0.15em}
\hspace{-1.5em}
\begin{subfigure}[b]{0.25\textwidth}
 \centering
 \includegraphics[width=\textwidth]{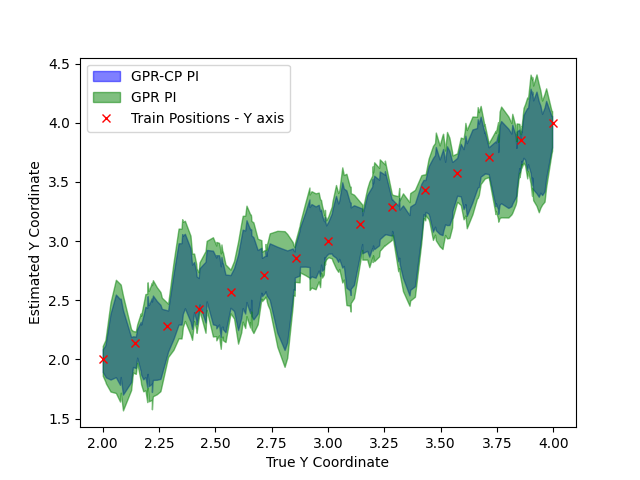}
 \vspace{-2em}
\end{subfigure}
\end{center}
\caption{\ac{PI} widths for $99\%$ coverage at $T_{60}=700$~ms and SNR$=15$~dB. Left: X-coordinate. Right: Y-coordinate.\looseness=-1}
\label{fig:pi_demo}
\vspace{-1.5em}
\end{figure}

Figure~\ref{fig:pi_demo} depicts the \acp{PI} produced for the X and Y axes as a function of the true position, with red x-marks denoting the labeled positions. \BL{For each axis, we varied the test positions uniformly across its range while keeping the other axis fixed at $3.0$ m.} 
It can be seen that GPR-CP often produces tighter intervals compared to GPR. In addition, the interval width is smaller near labeled samples, as expected.
\BL{Moreover, the width of the \acp{PI} for each axis is apparently influenced by the distribution of microphones. With only a single node positioned along the Y-axis, compared to four nodes along the X-axis, the predictions for the Y-axis are less certain, resulting in wider intervals.} \looseness=-1 

\section{Conclusions}
We propose a method for \ac{UQ} in estimating the speaker location in noisy and reverberant environments. Our approach utilizes a small amount of labeled data and additional unlabeled samples from unknown source locations. The core of our method is a \ac{GPR} model with a manifold-based kernel function, integrated with a \ac{CP} framework, tailored specifically for \ac{GPR}. This combination enables the construction of valid \acp{PI} with guaranteed coverage, which can be efficiently computed. Results are demonstrated across different simulated acoustic conditions.\looseness=-1 

\bibliographystyle{IEEEbib}
\bibliography{papers}

\end{document}